\documentclass[3p,times,twocolumn]{elsarticle}

\usepackage{ecrc}

\volume{00}

\firstpage{1}

\journalname{Nuclear Physics B Proceedings Supplement}

\runauth{Carsten Rott}

\jid{nuphbp}

\jnltitlelogo{Nuclear Physics B Proceedings Supplement}

\usepackage{amssymb}

\usepackage[figuresright]{rotating}

\begin{document}

\begin{frontmatter}



\dochead{}

\title{Review of Indirect WIMP Search Experiments}

\author[label1]{Carsten~Rott}
\address[label1]{Dept.~of Physics and Center for Cosmology and AstroParticle Physics,\\ The Ohio State University, Columbus, OH 43210, USA}

\begin{abstract}
Observational evidence for dark matter can be explained by Weakly Interacting Massive Particles (WIMPs). These dark matter particle candidates could indirectly be detected through the observation of signals produced as part of WIMP annihilations or decays. Latest results from indirect searches for WIMPs are reviewed. Current and planned experiments are presented and their prospects and discovery potential discussed.
\end{abstract}

\end{frontmatter}



\section{Introduction}

Despite overwhelming evidence that it composes the vast majority of the mass in the Universe, dark matter's particle properties literally remain in the dark. Identifying the mysterious nature of dark matter is one of today's most pressing scientific problems and is being sought for using colliders, direct-detection experiments, and powerful indirect detection techniques. WIMPs -- Weakly Interacting Massive Particles (denoted $\chi$ ) are attractive candidates for dark matter and naturally arise in many theories beyond the standard model of particle physics, which were developed to explain the origin of electroweak symmetry breaking and solve the gauge hierarchy problem~\cite{Bertone:2004pz}.

This review is structured in the following way. Evidence for dark matter and expected signals are discussed first, followed by a review of experiments and recent results. Prospects for the detection of WIMPs with existing and proposed experiments are discussed before concluding. 

\section{Evidence for dark matter and signals}
\label{sec1}
\subsection{Evidence for dark matter}

\begin{figure}
\begin{center}
\resizebox{\linewidth}{!}{\includegraphics{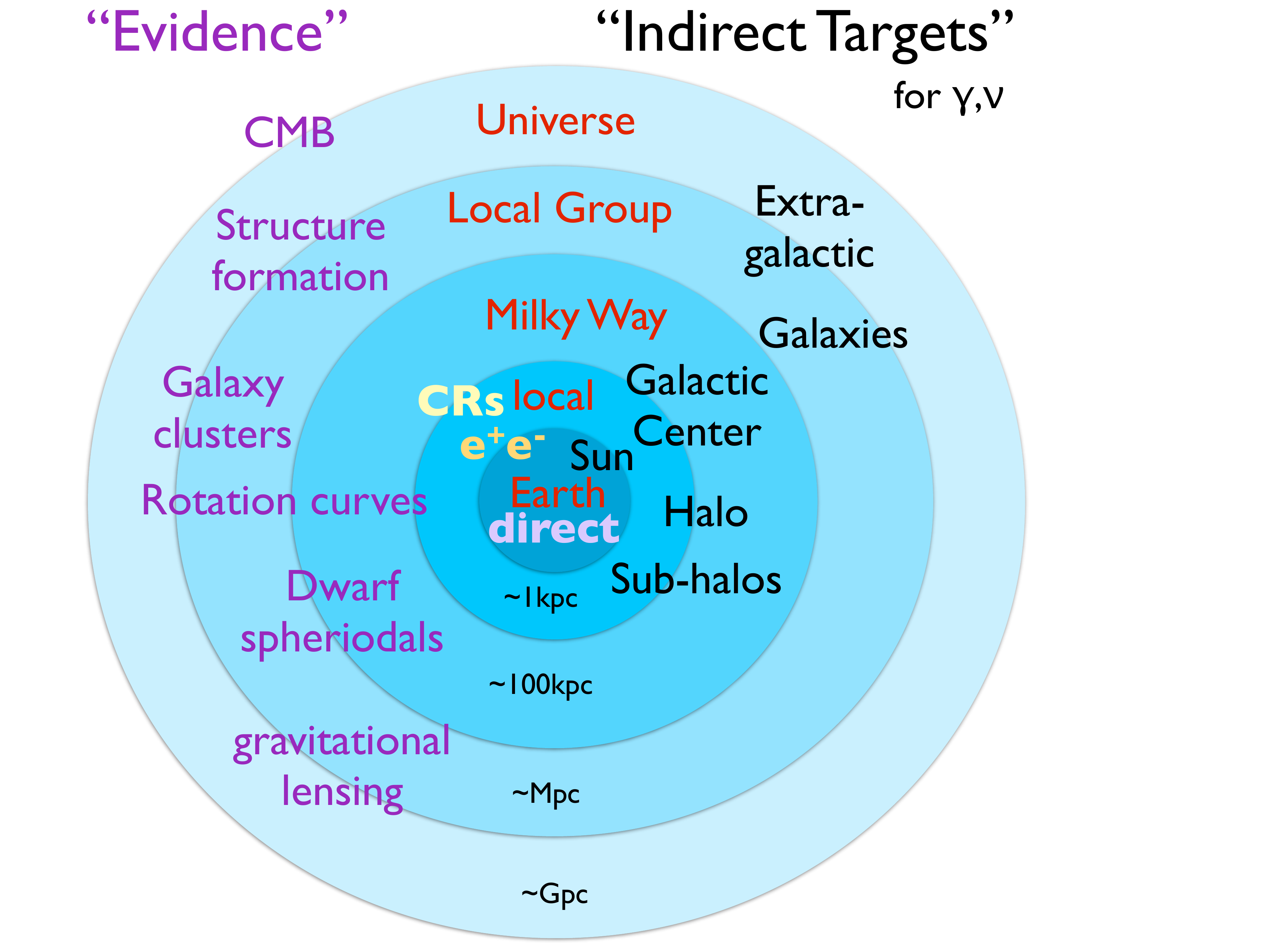}}
\caption{Observational evidence for dark matter and potential target that are expected contain significant amounts of dark matter.\label{fig:evidence}}
\end{center}
\end{figure}

Evidence for the existence of dark matter can be obtained at all scales from the motion of stars in dwarf spheroidal galaxies, galactic rotation curves,  virial velocities in clusters of galaxies, gravitationally lensed galaxies, baryon acoustic oscillations (BAO), to imprints on the cosmic microwave background (CMB). 
N-body simulations of dark matter can reproduce observed large scale structures in our universe and can also be used to obtain expected dark matter distributions in Milky Way like galaxies. Dark matter only simulations yield spherically symmetric halo density profiles $\rho(r)$, which describe the average dark matter density as function of the distance $r$ from the Galactic center. The impact of baryons, the behavior for small r (cusp-core problem), and the impact of sub-structure are still topics of debate.
Figure~\ref{fig:evidence} summarizes the observational evidence and lists astronomical objects that are expected to contain high dark matter densities and therefore are natural targets for indirect searches.

\subsection{Thermal relic}

If dark matter is a WIMP that is a thermal relic of the early Universe, then its total self-annihilation cross section averaged over the velocity distribution $\langle \sigma_A v \rangle$ is revealed by the observed dark matter abundance. Evolution of the number density $n$ is determined by the competition between production and annihilation with standard model particles.

\begin{equation}
\frac{dn}{dt} + 3 H n = \frac{d(na^3)}{a^3 dt} = \langle \sigma_A v \rangle \left( n^2_{eq} - n^2 \right)
\end{equation}

The Hubble parameter, $H$, provides a measure of the universal expansion rate and $n_{eq}$ is the equilibrium abundance $n_{eq}=g_{\chi}(mT/(2\pi))^{3/2}{\rm exp}(-m/T)$.  
In equilibrium $n$ decreases exponentially as the Universe expands and cools. When the $\chi$ abundance becomes very small, equilibrium can no longer be maintained and WIMPs freezes out~\cite{Steigman:2012nb}, setting the natural scale for $\langle \sigma_A v \rangle$.

\subsection{Indirect Detection}

Indirect signals from WIMP annihilations depend on the square of the dark matter density, $\rho$, the annihilation products, and propagation effects. As an example we discuss the differential neutrino flux from WIMP annihilations of mass $m_{\chi}$, which is given by:

\begin{equation}
\frac{d\Phi}{dE}(E,\phi,\theta) = \frac{1}{4 \pi} \frac{\langle \sigma_A v \rangle}{2 m_{\chi}^2}  \sum_f \frac{dN}{dE} B_f J(\Delta \Omega)
\end{equation}
where $\frac{dN}{dE}$ is the differential neutrino multiplicity per annihilation weighted by the branching fraction $B_f$. This annihilation part encodes the particle physics in form of the energy spectrum. Intrinsic WIMP properties could be derived from spectral shapes, as well as the WIMP mass identified via end-point analyses. The ``Astrophysics factor'' $J$ has a characteristic spacial dependence. $J$ is the integral over the line of sight (los) of the squared WIMP density over the integration solid angle $\Delta \Omega$, defined as
\begin{equation}
J(\Delta \Omega) = \int_{\Delta \Omega (\phi, \theta)} d\Omega' \int_{\rm los} \rho^2(r(l,\phi'))dl(r,\phi') 
\end{equation}
Note, that the dark matter annihilation cross sections could be velocity-dependent~\cite{Campbell:2010xc}. 

Expected fluxes of other cosmic messenger particles ($\gamma$-rays, $\bar{p}$/p, $e^{\pm}$, ...) can be obtained similarly, but often require extensive modeling of diffusion and propagation effects. WIMPs can be searched for with all these messengers, but $\gamma$-rays and neutrinos provide best detection prospects as spectral and directional information can be directly linked to the WIMP annihilations. Searches in the CMB are not discussed here~\cite{Finkbeiner:2011dx}.

\section{Instruments}

PAMELA -- Payload for Anti-Matter Exploration and Light-nuclei Astrophysics, is a satellite-borne magnetic spectrometer attached to the Resurs-DK1 satellite. Charged particles passing through the instrument aperture are deflected by the field of a neodymion-iron-boron permanent magnet. The resulting curvature of charged particle trajectories is measured precisely by a tracking system consisting of 6 plane double-sided silicon micro-strip tracker and depends on the particle charge and magnetic rigidity ($R=pc/|Z|e$). The energy and interaction topology of particles is measured with a sampling imaging calorimeter, in which pairs of orthogonal ministrip silicon sensor planes are interleaved with tungsten absorber plates. The main purpose of the calorimeter is to distinguish $e^{\pm}$ from $p/\bar{p}$ and He. Plastic scintillation counters act as an anti-coincidence system to reject particles that have entered through the sides of the instrument. Further detector components are a time of flight system, neutron calorimeter, and bottom scintillator~\cite{Picozza:2006nm}.

Fermi-LAT -- The Fermi Large Area Telescope is a pair-conversion telescope and the primary instrument on the Fermi Gamma-ray Space Telescope launched on June 11, 2008. It consists of three detector subsystems: (1) A tracker/converter with 18~layers of paired silicon strips detectors interleaved with tungsten foils, (2) An 8 layer CsI(Ti) scintillation crystals calorimeter, (3) plastic scintillator tiles and wavelength-shifting fibers as anti-coincidence detector. Fermi-LAT is sensitive in the energy range of $20~{\rm MeV} - 300~{\rm GeV}$. Fermi-LAT observations started in August 2008 and  first analyses were performed with the {\tt Pass 6} data event analysis scheme designed prior to launch. {\tt Pass 7} has been improved by accounting for known on-orbit effects. {\tt Pass 8} is still being developed, but it is redoing everything from the event reconstruction up~\cite{Fermi_Perform}. 

IACTs -- Imaging Atmospheric Cherenkov Telescopes (IACTs)~\cite{Funk:2012gq} detect the Cherenkov light emitted in atmospheric particle showers with duty cycles of about 15\% at a field of view (FOV) between $3.5^{\circ}$ to $5.0^{\circ}$ depending on the instrument. By using the atmosphere as a target their effective area is about a factor 500~larger than that of Fermi-LAT ($\sim1{\rm m}^2$).
VERITAS, operational since September 2007, is located at the Whipple Observatory Site (Southern Arizona) at an altitude of 1250~m. It provides sensitivity to a wide range of energies (150~GeV - 30~TeV) through stereoscopic imaging with four 12-meter telescopes. 
MAGIC~\cite{Aleksic:2011bx}, located at the Canary Island of La Palma at 2200~m, provides a low energy threshold of about 50~GeV with Stereo IACTs with two 17~m telescopes. Regular stereo observations were performed since Fall 2009.
The High Energy Stereoscopic System (H.E.S.S.) is located at Khomas Highland of Namibia at an altitude of 1800~m consists of an array of four 13-meter IACTs and achieves an energy threshold of about 200~GeV. Recently, the $600{\rm m}^2$ (28-meter-sized mirror) telescope H.E.S.S.~II has been added.

IceCube -- The IceCube neutrino telescope instruments a volume of about one gigaton of Antarctic ice beneath the surface of the Geographic South Pole with 5160 digital optical modules (DOMs) on 86 strings~\cite{see_Greg}. Each DOM contains a 10-inch PMT. IceCube exploits the good optical properties of the ice beneath the South Pole to detect neutrinos through the Cherenkov light emission of secondary particles produced by neutrino interactions.

Construction was completed after 6~years in December 2010. Data acquired during the construction phase has been analyzed.

ANTARES -- The ANTARES neutrino telescope~\cite{Aguilar:2010aa} is located between the depth of 2000~m -- 2475~m in the Mediterranean Sea, roughly~40~km offshore from Toulon in France. The full detector consists of twelve vertical lines equipped with a total of 885~10-inch PMTs (R7081-20 from Hamamatsu).

Super-K -- The Super-Kamiokande detector~\cite{Fukuda:2002uc} is the world's largest underground water Cherenkov
detector, located in Kamioka
mine (1000~m of rock overburden) in Hida-city, Gifu, Japan.
The cylindrical 50~kt with a diameter of 39.3~m and a height of 41.4~m began operating in April 1996.
Cherenkov light is used to reconstruct the position of neutrino interactions in the detector (contained events) and up-going muons passing through the detector, which originate from muon neutrinos interacting outside the detector volume.

\section{Summary of current results}

\subsection{Charged cosmic-rays}

WIMP annihilations in the Milky Way halo could yield positron, antiproton and antideuteron signals. The complex interplay of these messengers is discussed elsewhere (e.g.~\cite{Bottino:2000gc}) and we focus here on positrons.

\begin{figure}
\resizebox{\linewidth}{!}{\includegraphics{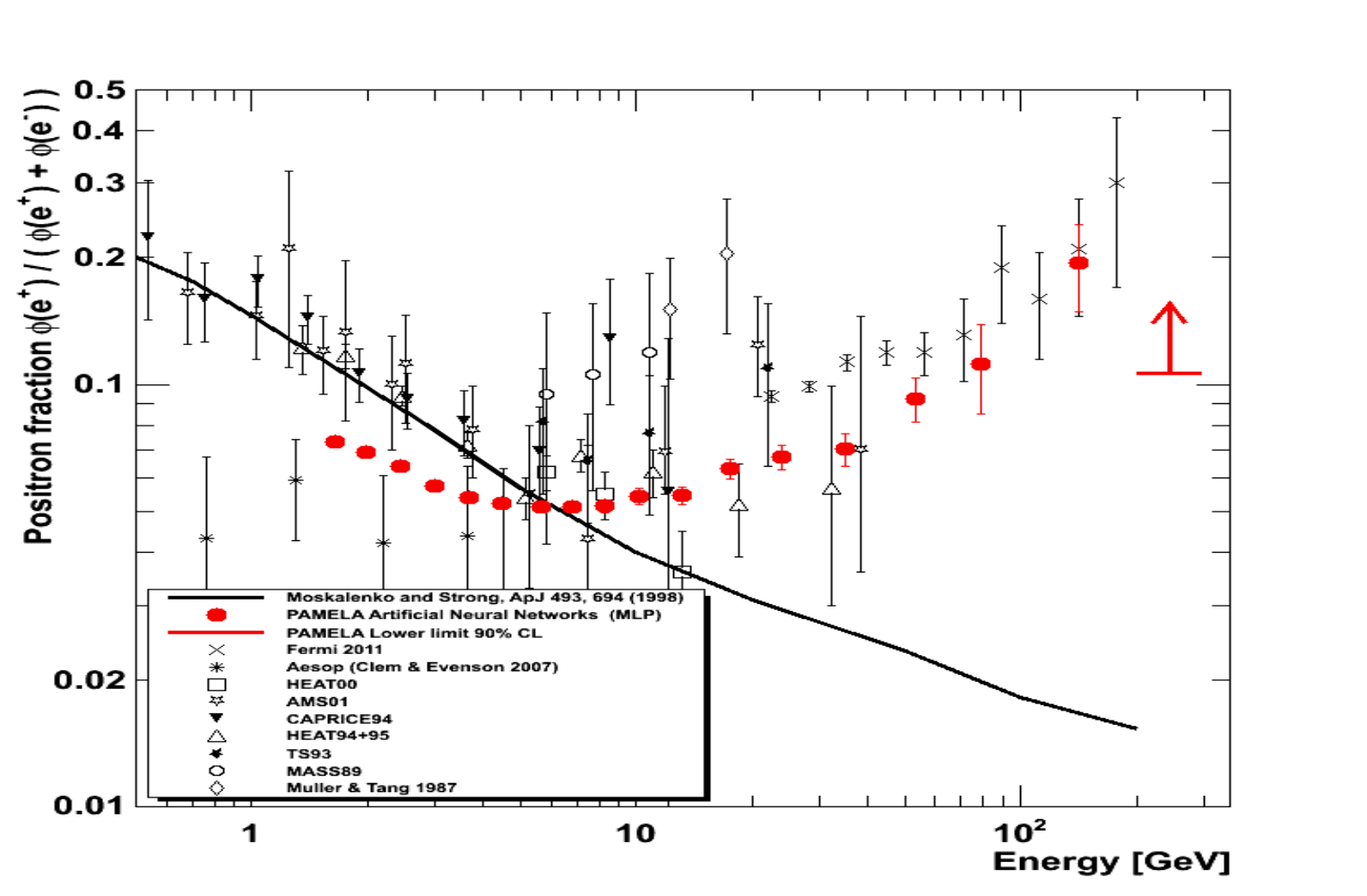}}
\caption{Positron fraction as function of the electron energy. PAMELA, HEAT, AMS-01, and Fermi-LAT data show a rising fraction above 10~GeV. The line indicates the expected positron fraction as obtained from standard electron-positron models.
\label{fig:PAMELA}}
\vspace{-1.\baselineskip}
\end{figure}

Measurements of the positron fraction $\phi(e^{+})/(\phi(e^{+})+\phi(e^{-}))$ show an unexplained rise above 10~GeV, that is inconsistent with standard secondary production models~\cite{Moskalenko:1997gh}. The excess, first observed by PAMELA~\cite{Adriani:2008zr} following initial hints by HEAT and AMS-01, has been confirmed by Fermi-LAT by back tracking leptons in the Earth's magnetic field. The observed positron anomaly is shown in Fig.~\ref{fig:PAMELA}~\cite{FermiLAT:2011ab}.

Due to the small gyroradius of electrons in the Galactic magnetic fields, electrons lose energy quickly. Hence, the source of the lepton excess must be of local nature and originate within about one kpc of the Sun. It could be due to unaccounted or limitations in the modeling of nearby astrophysical sources (e.g. pulsars~\cite{Yuksel:2008rf}, supernova remnants~\cite{Blasi:2009hv}), or be a first hint of a dark matter signal.
If the positron excess in combination with the energy spectra of electrons and positrons above 300~GeV measured by Fermi~\cite{Abdo:2009zk}, H.E.S.S.~\cite{Aharonian:2009ah}, and ATIC~\cite{Chang:2008aa} are interpreted as originating from WIMP annihilations or decays, then it indicates leptophilic models that require enhancement in annihilation rate compared to the thermal relic cross section. Such scenarios are presently disfavored by $\gamma$-ray, radio, and neutrino observations.

\subsection{Gamma-rays / radio}

Combined Fermi-LAT observations of various dwarf spheroidal galaxies have constrained WIMP self-annihilations with the thermal relic cross section for WIMP masses below $10$~GeV~\cite{Ackermann:2011wa}, limits significantly weaken above on TeV. MAGIC has investigated dwarfs Willman~1~\cite{Aliu:2008ny} and Segue~1 and obtained limits around $10^{-22} {\rm cm}^3{\rm s}^{-1}$ and $10^{-23} {\rm cm}^3{\rm s}^{-1}$ for annihilation in $\mu^{+}\mu^{-}$ and $\tau^{+}\tau^{-}$, respectively, for WIMP masses between 200~GeV and 1.2~TeV~\cite{Aleksic:2011jx}.
VERITAS observed no significant gamma-ray excess from Segue1~\cite{Aliu:2012ga} and four dwarf spheroidals and derived an upper limit on the gamma-ray flux to constrain the self-annihilation cross section.  The most stringent limit of $10^{-22} {\rm cm}^3{\rm s}^{-1}$ is obtained for $m_{\chi} = 300$~GeV~\cite{Acciari:2010ab}.
H.E.S.S. reported cross section constraints from $10^{-21}{\rm cm}^3{\rm s}^{-1}$ to $10^{-22}{\rm cm}^3{\rm s}^{-1}$ from the Sculptor and Carina dwarf galaxies depending on the assumed halo model~\cite{Abramowski:2010aa}. Limits can also be obtained from observations of $\gamma-$rays, produced via final state radiation, inverse Compton scattering, or synchrotron radiation, of the Galactic Center and Galactic Ridge regions, as well as radio observation of the Galactic Center~\cite{Meade:2009iu}. Cosmic X-ray data further provides constraints on WIMP annihilation~\cite{Zavala:2011tt}. 

Spectral features such as a spectral line can be a convincing signature for WIMP annihilations and could be produced in $\chi \chi \rightarrow \gamma \gamma$, $h \gamma$, and $Z \gamma$. In a search by the Fermi-LAT collaboration in 2~yrs of {\tt Pass 6} data, no signal was observed in the ROI (region of interest), which consisted of the entire sky, but excluded the Galactic plane and known $\gamma$-ray sources (1FGL). Assuming the background is described by a power law with a spectral index free to vary, limits were derived on $\langle \sigma_{A} v \rangle$~\cite{Ackermann:2012qk}. The obtained limits are in mild tension with claims of an observation of a 130~GeV $\gamma$-ray line near the Galactic center in 4~years of {\tt Pass 7} data~\cite{Weniger:2012tx,Bringmann:2012vr}. While it is unlikely that the line originates from astro-physical background~\cite{Aharonian:2012cs}, it could be an instrumental effect, for example as the result of a non-linear energy mapping. Fortunately, the energy mapping can be tested in Earth limb data. While the Fermi-LAT collaboration will have to clarify this topic, initial studies show indications of non-linear effects~\cite{Finkbeiner:2012ez}. H.E.S.S.~II data or radio data could eventually also be used confirm or refute the line in independent measurements. For annihilations with final states $Z \gamma$ or $h \gamma$, relativistic $e^{\pm}$ are expected, which would generate  synchrotron radiation when interacting with Galactic magnetic fields. Existing radio data in the Galactic Center is already in minor tension with the presence for a contracted NFW profile. Currently running and future radio telescopes like Long Wavelength Array, LOFAR and SKA have great potential in resolving this question further~\cite{Fornengo:2011iq,Laha:2012fg}.

\subsection{Neutrinos}

\begin{figure}[h]
\resizebox{\linewidth}{!}{\includegraphics{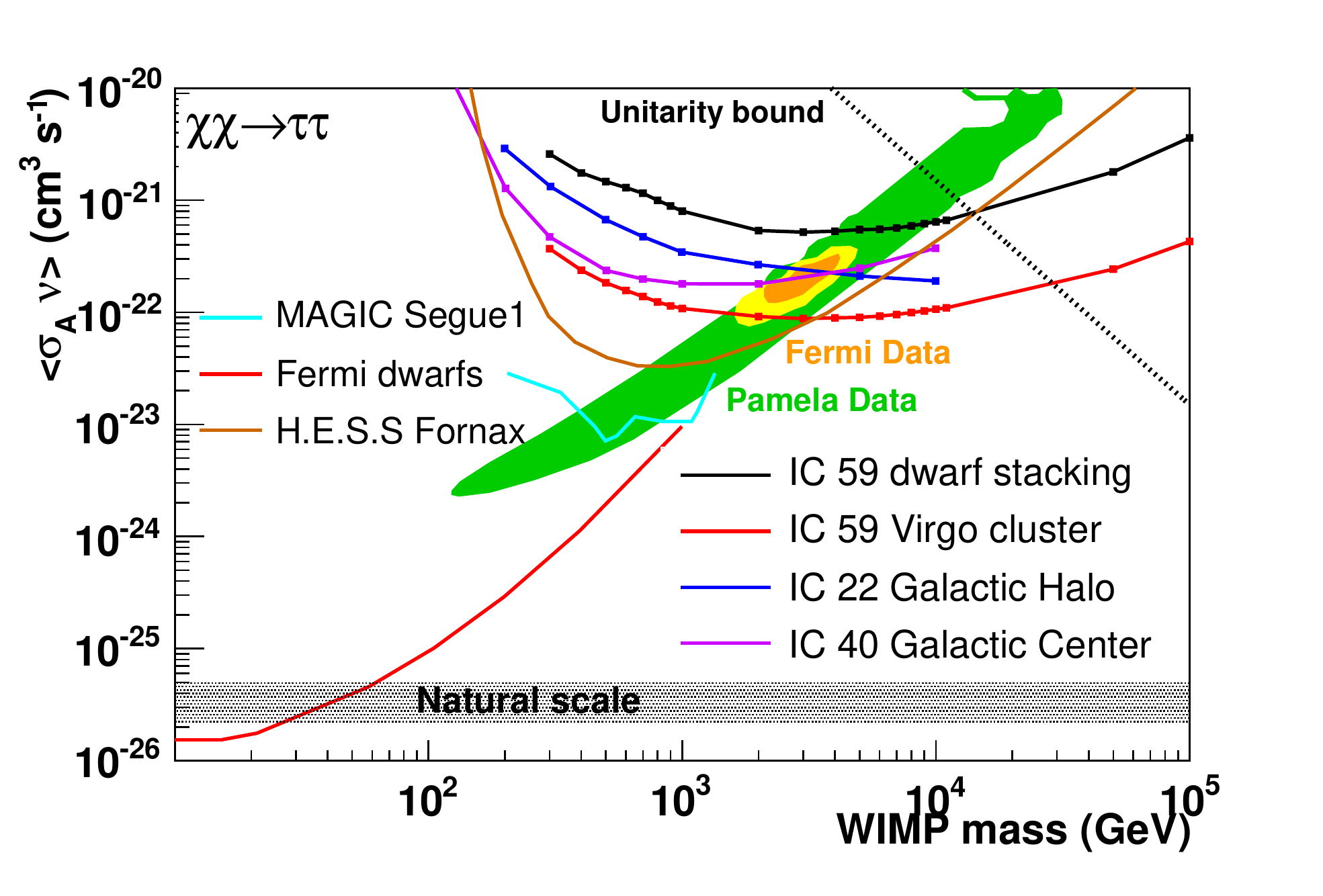}}
\caption{Comparison of gamma-ray bounds with those from neutrino telescopes and a dark matter model motivated by the PAMELA positron excess for $\chi\chi \rightarrow \tau^{+}\tau^{-}$~\cite{Meade:2009iu}.
\label{fig:gamma_nu}}
\vspace{-1.\baselineskip}
\end{figure}

The indirect search for WIMP-induced neutrinos aims to detect Galactic signals similar to the searches with gamma-rays, but also from self-annihilating WIMPs captured by the Sun and Earth.
IceCube has searched for signals from the Galactic halo~\cite{Abbasi:2011eq} and Galactic center~\cite{IceCube:2012ws} and improved upon theoretical predictions~\cite{Yuksel:2007ac}. Tight constraints were also derived from dwarf spheroidal galaxies~\cite{IceCube:2011ae} and the Virgo cluster. Figure~\ref{fig:gamma_nu} shows a comparison of these present bounds on the dark matter self-annihilation cross section as function of the WIMP mass for neutrinos and gamma-rays. Neutrinos are in particular competitive for $m_{\chi}>1$~TeV, with best sensitivity achieved by clusters of galaxies if substructure is taken into account~\cite{Dasgupta:2012bd}.

\begin{figure}[h]
\resizebox{\linewidth}{!}{\includegraphics{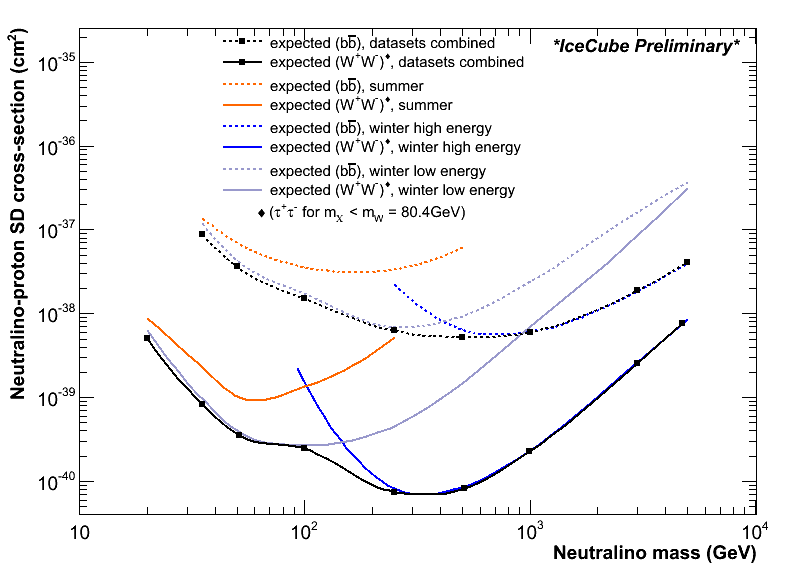}}
\caption{IceCube 79-string detector solar WIMP sensitivity~\cite{matthias}.
\label{fig:SolarWIMP_sens}}
\vspace{-1.\baselineskip}
\end{figure}

WIMPs could accumulate in the Sun or Earth and give rise to detectable neutrino signals. Energy loss induced by an initial scatter of a WIMP on a nucleon in the Sun can lead to the gravitational capture in the Sun. The probability for such an interaction, which is the same underlying physics process as being searched for in direct detection experiments, depends on the WIMP nucleon scattering cross section. WIMPs accumulate in the Sun and start annihilating. The annihilation rate depends on the amount of dark matter in the Sun. The annihilation rate steadily increases with the number of thermalized WIMPs near the center of the Sun up to a point where it becomes equal to the capture rate. At this point equilibrium has been reached and the annihilation rate is independent of the self-annihilation cross section. The neutrino flux from the Sun hence depends only on the capture rate, which can then be linked to the WIMP nucleon scattering cross section. As the Sun is primary a proton target, in particular tight constraints can be derived on the spin-dependent WIMP-proton scattering cross section $\sigma_{\chi p}$. Figure~\ref{fig:SolarWIMP_sens} shows IceCube's sensitivity with one year of data collected with the 79-string detector. The dataset has been divided in three independent categories (summer, winter low-energy, winter high-energy) and is later combined.

\begin{figure}
\resizebox{\linewidth}{!}{\includegraphics{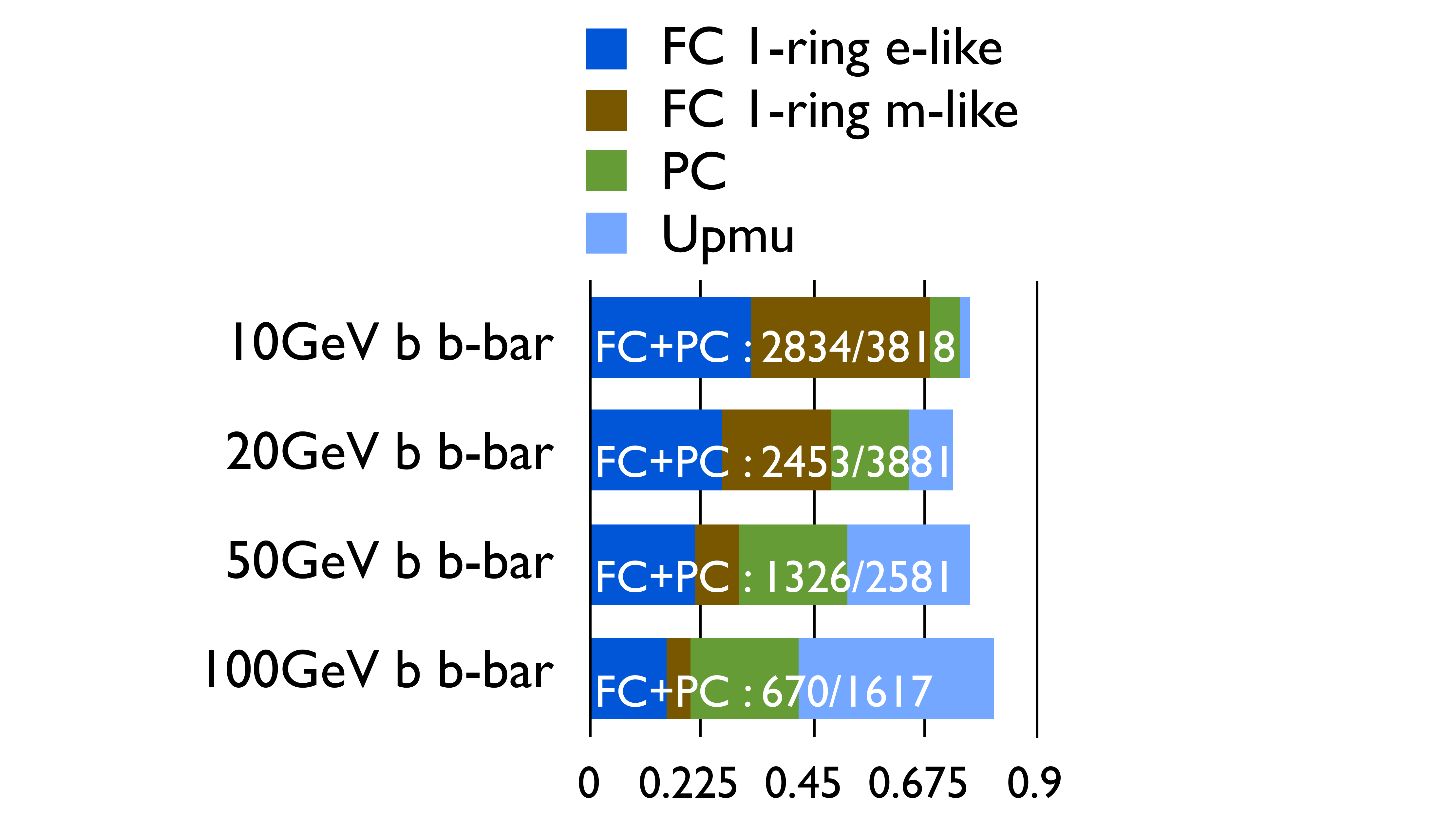}}
\caption{Super-K solar WIMP analysis: Distribution among fully contained (FC), partially contained (PC), and up-going muons (upmu) events expected at final analysis level for a WIMP of given mass annihilating into $b\bar{b}$.  For WIMPs masses below 50~GeV the dominant signal is expected from contained events (FC+PC)~\cite{koun}.
\label{fig:SolarWIMP_SK}}
\vspace{-1.\baselineskip}
\end{figure}

\begin{figure}[h]
\resizebox{\linewidth}{!}{\includegraphics{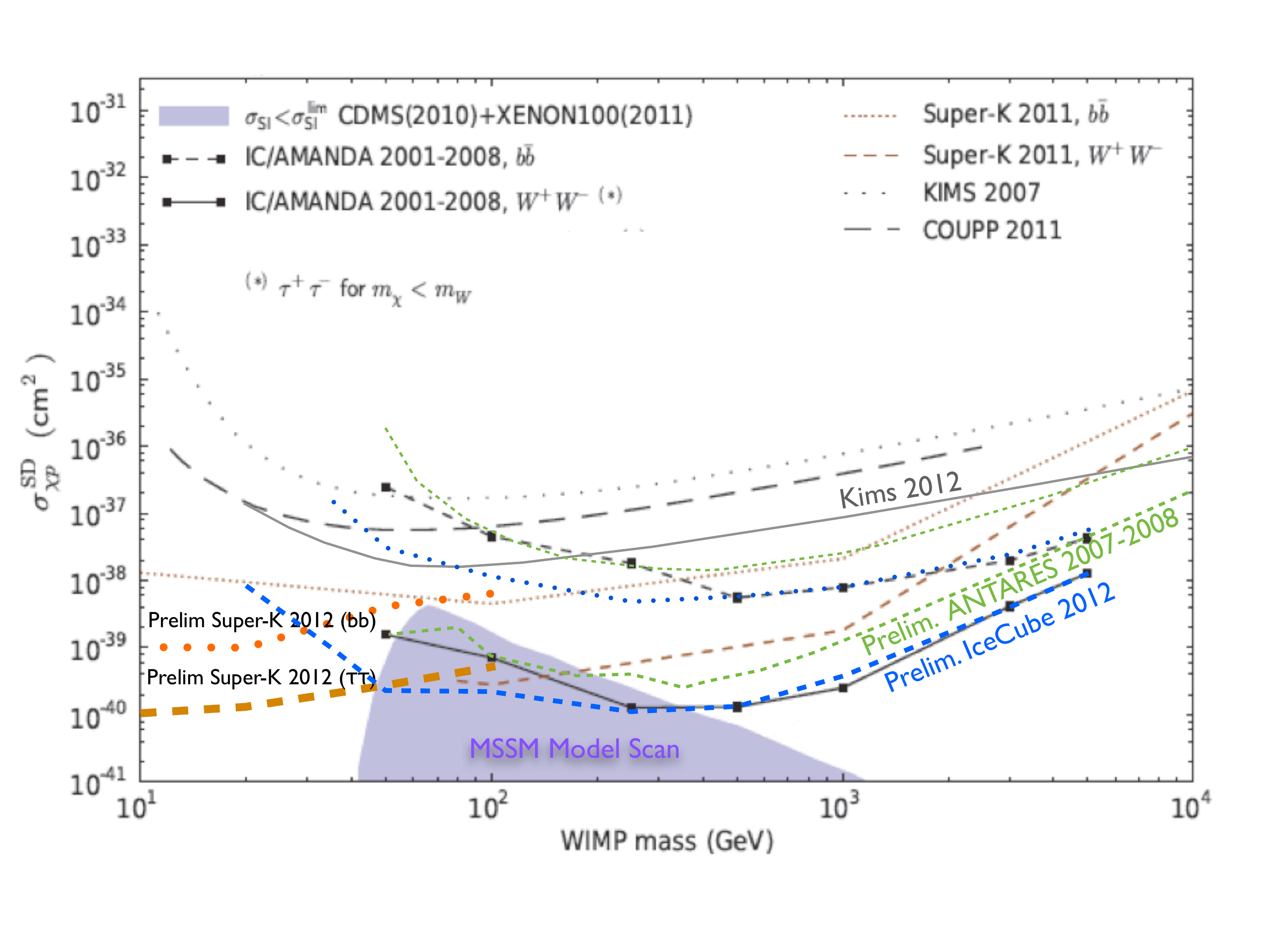}}
\caption{Constraints on the spin-dependent WIMP-proton scattering cross section.
\label{fig:SolarWIMP}}
\vspace{-1.\baselineskip}
\end{figure}

Super-K searched the up-going muon sample for signals in 3109~days and derived limits~\cite{Tanaka:2011uf}, that were now improved upon after including fully and partially contained events (see Fig.~\ref{fig:SolarWIMP_SK}), which are of importance for low-mass WIMPs~\cite{Rott:2011fh}. The preliminary limits from Super-K are compared to those from ANTARES and IceCube in Fig.~\ref{fig:SolarWIMP}.

Limits on the WIMP-nucleon scattering cross section can also be deduced from limits on mono-jet and mono-photon signals at hadron colliders, however, they depend strongly on the choice of the underlying effective theory and mediator masses~\cite{Bai:2010hh}.

\section{Path Towards a Large Detector -- PINGU}

\begin{figure}[h]
\resizebox{\linewidth}{!}{\includegraphics{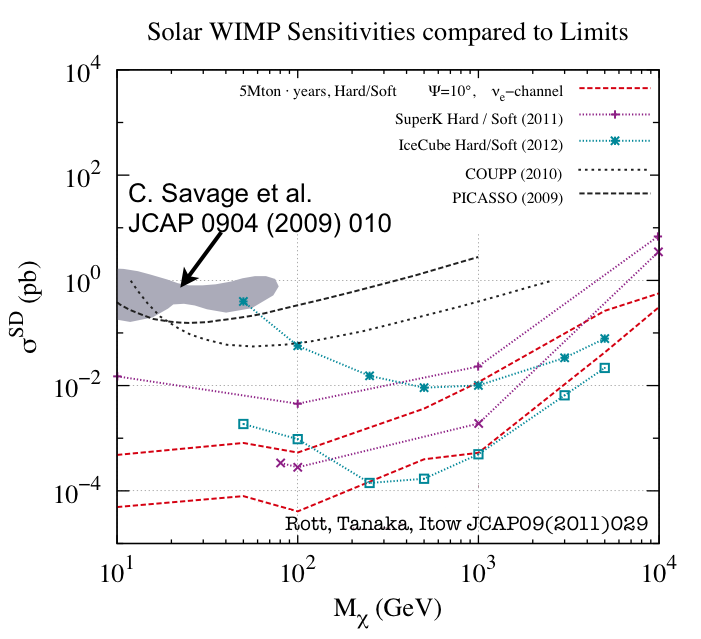}}
\caption{Sensitivity of a next generation neutrino detector assuming 5~Megaton years of data~\cite{Rott:2011fh}.
\label{fig:SolarWIMP_5Mton}}
\vspace{-1.\baselineskip}
\end{figure}

A conclusive test of many low-mass dark matter scenarios, a more precise study of atmospheric oscillation parameters, and an enhanced sensitivity towards supernova burst neutrinos would require a very large neutrino detector with a low energy threshold. Such a detector could be constructed in two phases at the geographic South Pole, making use of the existing infrastructure, good optical properties of the naturally occurring detector medium and support structure, and benefit from the IceCube detector to veto atmospheric muons.

A mega-ton-sized ring-imaging detector could be constructed in two stages. The first stage (PINGU -- Precision IceCube Next Generation Upgrade) would consist of an upgrade to the IceCube--DeepCore detector~\cite{Collaboration:2011ym} using existing technology complemented with calibration devices and novel optical modules. Physics results are guaranteed by relying on proven IceCube sensors, while the performance of new technologies towards stage~2 can be evaluated. For stage~1 it is envisioned to deploy about 20~string during two seasons. The aim is to achieve an energy threshold of a few GeV for this multi-mega-ton detector~\cite{DeYoung:2011ke,Karle:2012up}.
A stage~2 detector (MICA - Mega-ton Ice Cherenkov Array), consisting of the order of 100~strings, using a technology choice based on the performance of the stage~1 array would then aim at constructing a large ring-imaging detector. It is envisioned to reconstruct individual events above a threshold on the order of 100~MeV and use collective event information to detect supernova burst neutrinos. Primary physics motivations are dark matter searches, neutrino oscillation studies, and increased sensitivity towards supernova burst neutrinos. Extensions reaching proton decay could possibly be contemplated. Other physics topics include, but are not limited to: Sterile neutrinos and neutrinos from Galactic sources. 


The deep ice at the geographic South Pole possesses good optical properties below 2100~m and a high radio-purity. The absorption length at 400~nm is about $\lambda_{\rm abs}\approx 155$~m and effective scattering length is on the order of $\lambda_{\rm scat}^{\rm eff} \approx 47$~m. Uranium ($^{238}{\rm U}$) and Thorium ($^{232}{\rm Th}$) contaminations are very low at $10^{-4}$~ppb and Potassium at 0.1~ppb in the Antarctic ice~\cite{Cherwinka:2011ij}. The combination of low installation costs and the ability to build a contiguous detector, makes the South Pole an ideal site. However, the maximum density of instrumentation is determined by the installation procedure and will ultimately determine what photo coverage can be achieved. While the stage~1 detector can rely on the existing IceCube hot-water drilling technology for the stage~2 modifications are likely necessary. Drilling and deployment costs are expected to be below 10\% of the total costs of the array, making the ``excavation cost'' component a moderate one.

IceCube digital optical modules (DOMs)~\cite{Abbasi:2010vc} are functioning extremely well, which is underlined by the fact that the number of DOMs that fail commissioning is at a percent level and the number of lost DOMs after successful freeze-in and commissioning is a fraction of a percent. The IceCube detector is operating very stable and shows detector uptimes of about 99\%. DeepCore utilizes 252~mm diameter Hamamatsu R7081MOD (super bialkali photocathodes), which are identical to the standard IceCube PMTs (R7081-02), but with a quantum efficiency that is increased by 40\% at $\lambda=390$~nm. While, the physics goals of the stage~1 detector are achievable with the existing DeepCore sensors, it is intended to utilize also new photon detection technologies, with the goal to demonstrate the potential for reconstructing Cherenkov ring fragments. Developed for KM3NeT~\cite{KM3NeT}, multi-PMT optical modules, could be adapted for the use in the ice. Other optical devices utilizing wavelength shifter techniques to increase the photo sensitive area in a cost effective manner are also under consideration.

\section{Future prospects and outlook}
\subsection{Gamma-rays}
\begin{figure}[h]
\resizebox{\linewidth}{!}{\includegraphics{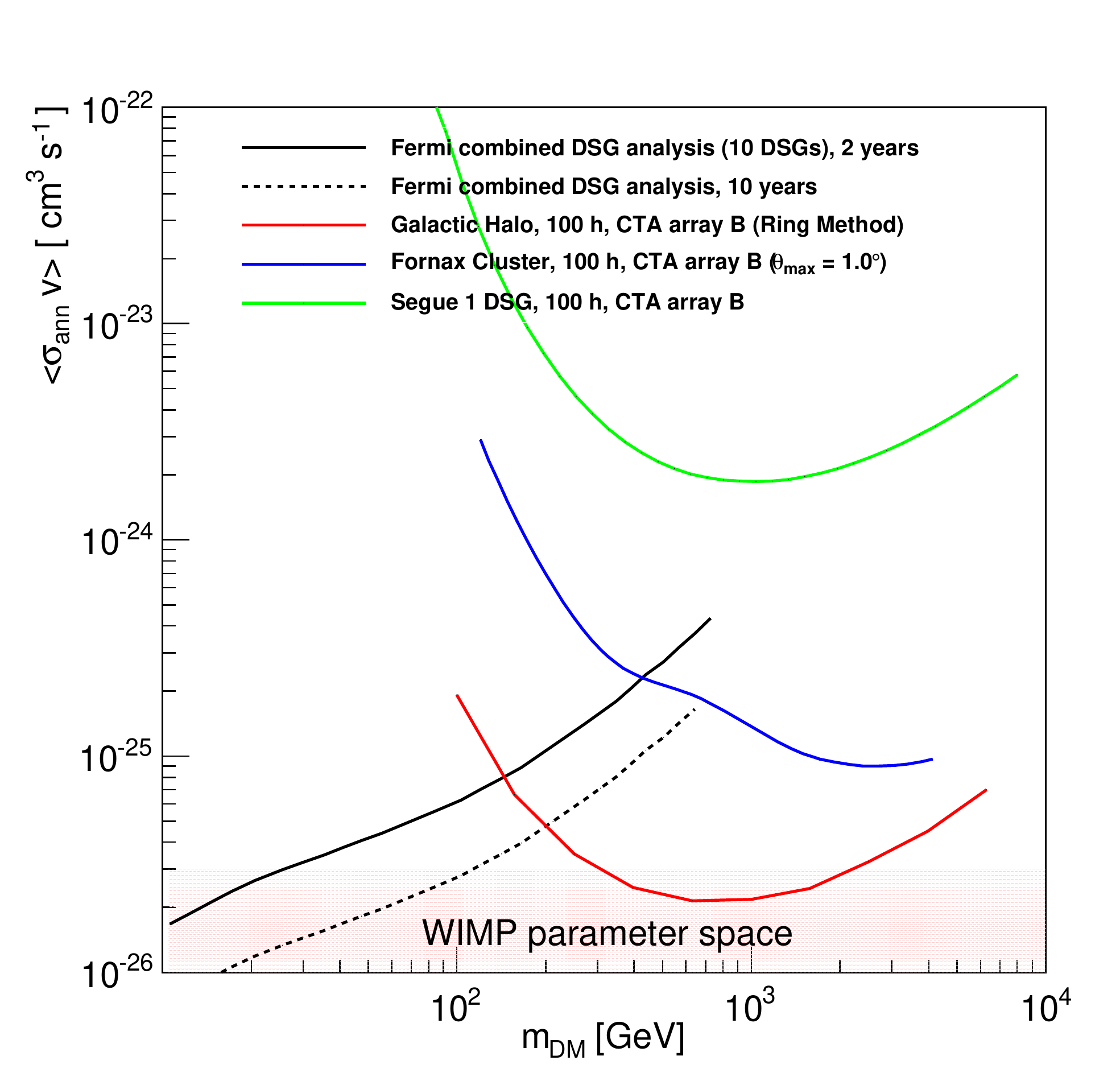}}
\caption{Comparison CTA sensitivity~\cite{Doro:2012xx} to current (2~years) Fermi-LAT exclusion limit~\cite{Ackermann:2011wa} and scaled to the 10~years expectation. 
\label{fig:gamma_future}}
\vspace{-1.\baselineskip}
\end{figure}

Fermi-LAT has produced already extremely competitive bounds on $\langle \sigma_{A} v \rangle$, future analyses will benefit from larger statistics and potential discoveries of new ultra-faint dwarfs. Figure~\ref{fig:gamma_future} shows a comparison of the Fermi-LAT dwarf spheroidal analysis~\cite{Ackermann:2011wa} compared to the expected sensitivity of the Cherenkov Telescope Array (CTA)~\cite{Doro:2012xx}, which is currently in the design phase. In the future GAMMA-400~\cite{Galper:2012fp} with an effective area of $4{\rm m}^2$, an angular resolution of $\sim 0.01^{\circ}$ at $E_{\gamma}=$100~GeV and an energy resolution of 1\% could surpass Fermi-LAT. GAMMA-400 could be launched as early as 2018.

\subsection{Neutrinos}

\begin{figure}[t]
\resizebox{\linewidth}{!}{\includegraphics{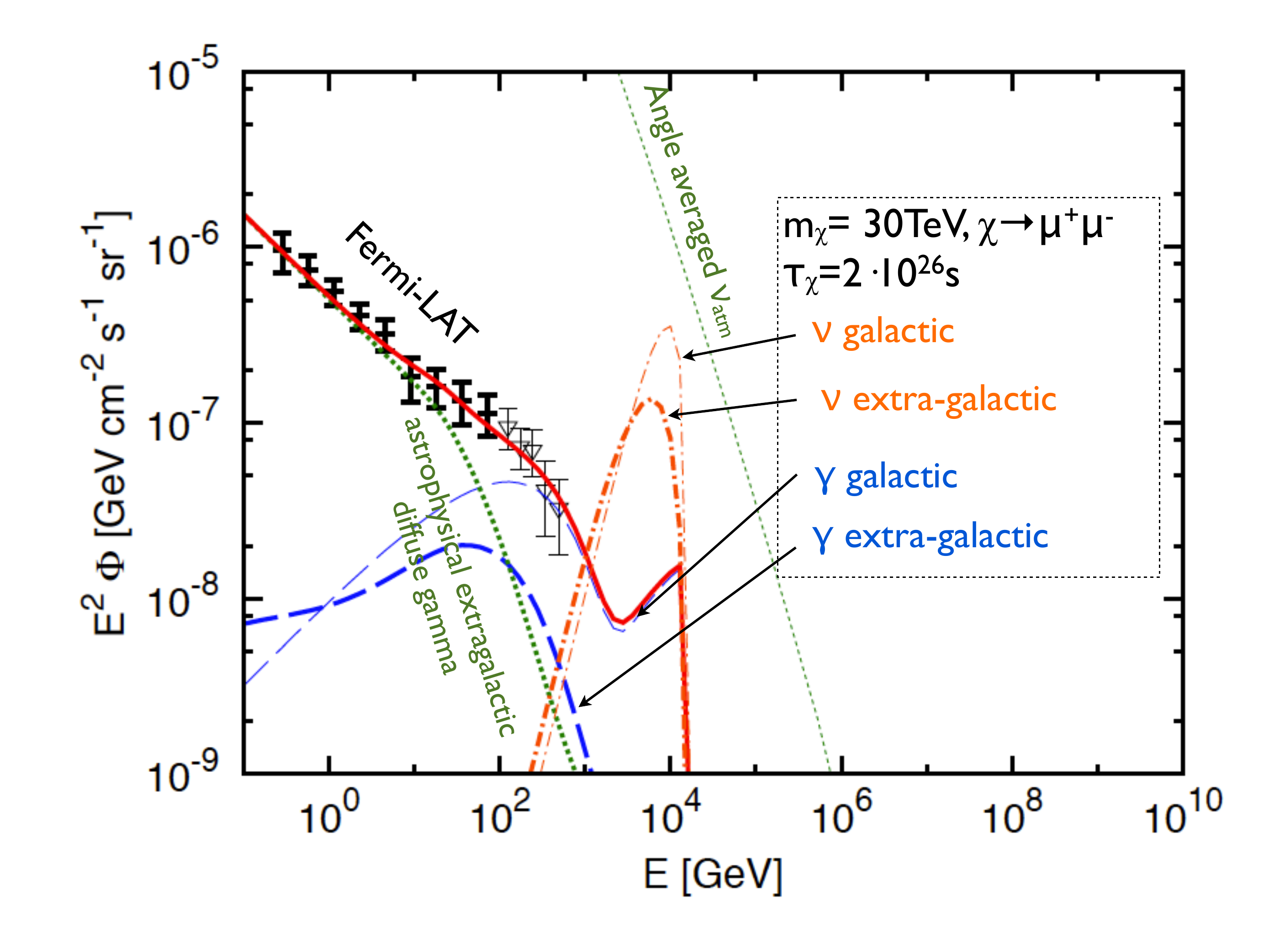}}
\caption{Extra-galactic gamma-ray flux as observed by Fermi-LAT and predictions for neutrino fluxes testable by IceCube~\cite{Murase:2012xs}.
\label{fig:Murase}}
\vspace{-1.\baselineskip}
\end{figure}

IceCube bounds presented at this conference only used the partially instrumented detector. More than a year of high-quality physics data has already been collected with the full IceCube detector, including 7 more strings (2 of them for DeepCore), further lowering its threshold. This data could also be used to test un-explained features in the extra-galactic gamma-ray fluxes observed by Fermi-LAT, that could hint at high mass dark matter~\cite{Murase:2012xs} (see Fig.~\ref{fig:Murase}).

Exciting prospect for dark matter searches also exist at next-generation neutrino detectors such as Hyper-Kamiokande~\cite{Abe:2011ts}, LENA~\cite{Wurm:2011zn}, and PINGU. 
WIMP scenarios motivated by DAMAs annual modulation signal~\cite{Savage:2008er} and isospin-violating scenarios~\cite{Feng:2011vu} motivated by DAMA and CoGeNT signals would be testable at ($\sim 1$GeV) - threshold neutrino detectors~\cite{Rott:2011fh} (see Fig.~\ref{fig:SolarWIMP_5Mton}). Further new detection channels such as low-energy neutrinos originating from hadronic particle showers created in WIMP annihilations in the Sun could enhances sensitivities and provide new ways to search for dark matter.

\subsection{Low-energy neutrinos}

\begin{figure}[t]
\resizebox{\linewidth}{!}{\includegraphics{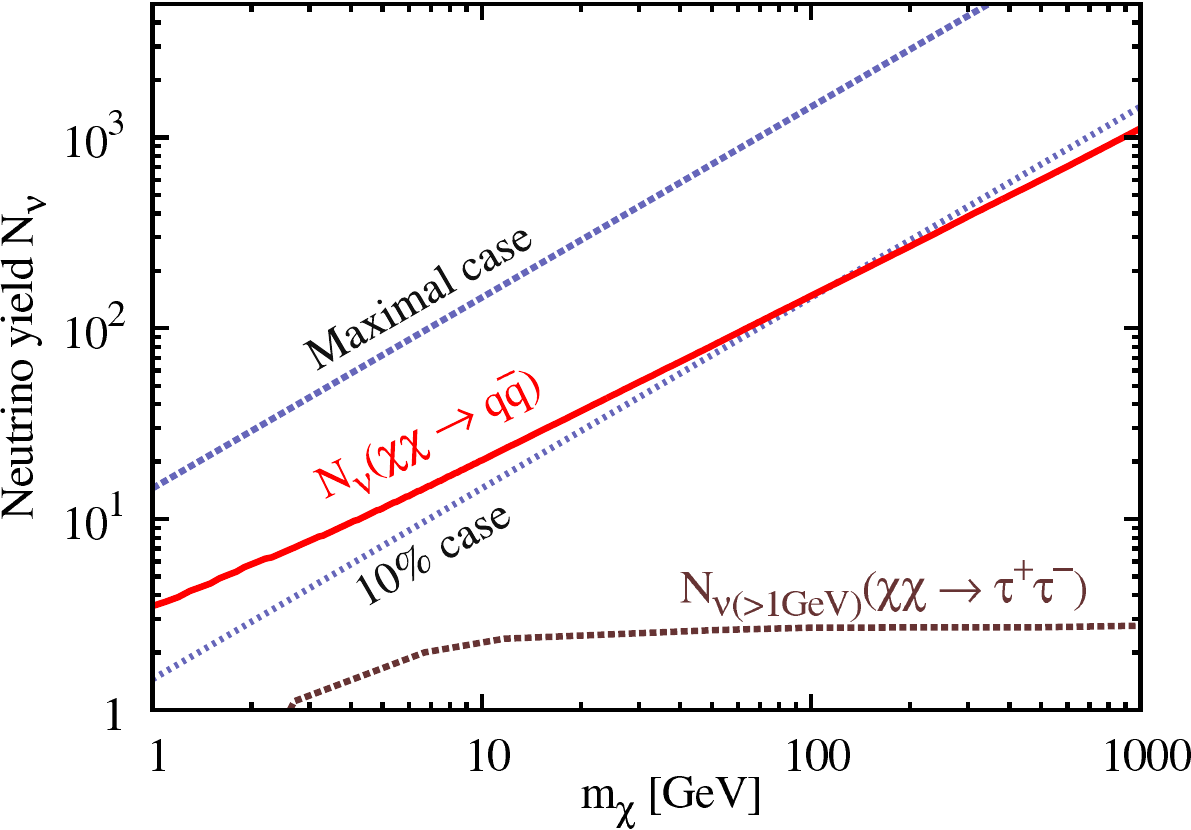}}
\caption{Low-energy neutrino yield (summed over flavors) per WIMP annihilation ($\sqrt{s}=2 m_\chi$) in the solar core, obtained by simulating pion-induced hadronic showers in the solar medium is compared to the high-energy neutrino yield for $\tau^{+}\tau^{-}$~\cite{Rott:2012qb}.  
\label{fig:pionyields}}
\vspace{-1.\baselineskip}
\end{figure}

\begin{figure}
\resizebox{\linewidth}{!}{\includegraphics{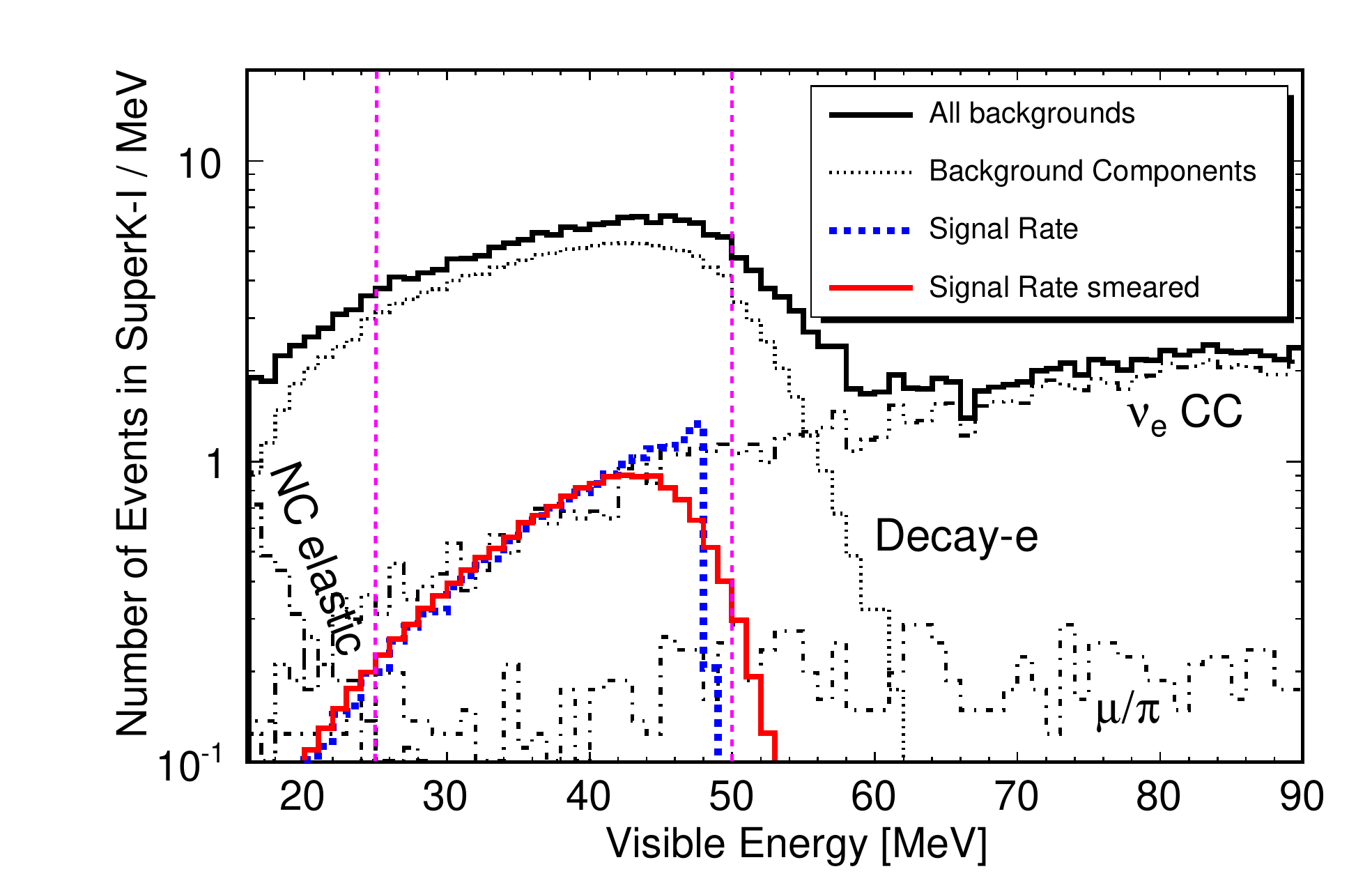}}
\caption{Example of a low-energy $\bar{\nu}_e$ signal from solar WIMP annihilation in Super-K, along with the measured background (4.1 livetime years).  The signal shape is independent of WIMP properties, and its normalization scales with $\sigma_{\chi p}^{\rm SD}$ (here chosen to be at the edge of exclusion)~\cite{Rott:2012qb}.
\label{fig:lowEsignal}}
\vspace{-1.\baselineskip}
\end{figure}

The large existing Super-K dataset and that of proposed future low-energy threshold neutrino detectors is sensitive to a new WIMP detection channel~\cite{Rott:2012qb,Bernal:2012qh}. A large number of low-energy neutrinos are expected from any annihilation with hadronic components in its final states in the Sun. Hadrons interact in the dense solar medium, producing a hadronic shower, which will result in a large number of pions (the pion multiplication effect is shown in Fig.~\ref{fig:pionyields}). Positive pions and muons decay at rest, producing low-energy neutrinos with known spectra, including $\bar{\nu}_e$ through neutrino mixing. A signal could be detected by Super-K (see Fig.~\ref{fig:lowEsignal}) and in the future provide a new probe of WIMP-nucleon scattering. Compared to other methods, the sensitivity is competitive and the uncertainties are complementary.

\section{Conclusions}

Indirect searches have resulted in tight constraints on fundamental properties of WIMPs. Neutrinos are highly competitive for extended (Galactic halo/center, Clusters of Galaxies, ...) and point-like sources (Dwarf spheroidal, ...) for energies above about 1~TeV. Gamma-rays are more sensitive in the detection of lower-mass WIMPs. Any detection of WIMPs would likely require an independent observation with both messengers.

WIMP annihilations in the Sun provide a discovery channel for neutrinos through a striking signature. Due to the opacity of the Sun to high-energy neutrinos originating in the center, Solar WIMP signals are detected in the energy range below 100~GeV. Neutrino detectors (IceCube and Super-K) provide the world's best limits and their sensitivity continues to improve as more data is collected and new detection channels are investigated.


\section*{Acknowledgments}

I would like to thank John~F.~Beacom, Basudeb Dasgupta, Shunsaku Horiuchi, Alexander Kappes, Matthew~Kistler, Ranjan Laha, and Kohta Murase for discussions and comments and Koun Choi, Jan Conrad, Yoshitaka Itow, and Piotr Mijakowski for providing materials. 


\end{document}